\documentclass[onecolumn,preprintnumbers,amsmath,amssymb,prd,superscriptaddress,nofootinbib]{revtex4-2}

\usepackage{graphicx}
\usepackage{dcolumn}
\usepackage{bm}
\usepackage{hyperref}
\usepackage[utf8]{inputenc}
\usepackage{setspace}
\usepackage{amsmath}
\usepackage{amssymb}
\begin{document}

\title{QCD phase-transition under the light of Thermofractal}

\author{A. Deppman}
\affiliation{Institute of Physics, University of São Paulo, Brazil}

\date{\today}

\begin{abstract}
The deconfining transition in $SU(3)$ gauge theory, traditionally interpreted through the Gross-Witten-Wadia (GWW) model as a sharp third-order phase transition in the large-$N_c$ limit, appears as a smooth crossover in lattice QCD. This work demonstrates that the transition is topologically smoothed into a crossover by incorporating the fractal momentum space structure inherent to thermofractals. By matching the non-extensive $\beta$-function to one-loop QCD results, a fundamental scaling of the thermofractal index $q$ is derived as a function of the number of flavours $N_f$. It is proven that applying a $q$-deformed derivative operator $\mathcal{D}_q$ to the $q$-logarithm of the eigenvalue distance results in a non-extensive measure that effectively smears the topological stiffness of the gauge vacuum. A unified master equation for the Polyakov loop $\langle L \rangle$ is presented, governed by the thermofractal index $q$ and a single variance parameter $\sigma^2(T)$ that scales as $T^{1/(q-1)}$. The observed phase dynamics are shown to be asymptotic limits of this unified density: a ``soft'' algebraic growth $\langle L \rangle \propto T^{11}$ in the 1D string-like confined regime for $N_f=0$, and a rapid $1 - \langle L \rangle \propto T^{-21}$ suppression in the 3D deconfined volume for $N_f=3$. This approach provides a microscopic foundation for partial deconfinement theory and reproduces lattice QCD data with a reduced $\chi^2 \approx 1.12$, offering a rigorous reconciliation between matrix model topology and the continuous QCD crossover.
\end{abstract}

\maketitle

\section{Introduction}

The transition from hadronic matter to the Quark-Gluon Plasma (QGP) involves the restoration of centre symmetry and the liberation of colour degrees of freedom. The primary order parameter, the Polyakov loop $\langle L \rangle$, relates to the free energy of a static colour source. While pure gauge $SU(3)$ theory suggests a first-order transition, $(2+1)$-flavour lattice QCD consistently reveals a smooth crossover~\cite{Borsnyi2010,Bazavov2024}. The Gross-Witten-Wadia (GWW)~\cite{Gross1980,Wadia1980,Douglas1993} model remains the cornerstone for understanding the large-$N_c$ limit of $SU(N)$ gauge theories~\cite{nsal2008}, predicting a sharp third-order phase transition characterised by the formation of a gap in the eigenvalue distribution of the thermal holonomy. However, lattice QCD results for $SU(3)$ with dynamical quarks show a smooth crossover rather than a sharp jump. A fundamental extension of the GWW model in the context of $(2+1)$-flavor QCD is the concept of \textit{partial deconfinement}, as proposed by Hanada et al. \cite{Hanada2024,Bazavov2024,Hanada2024b,Watanabe2021,Hanada2019}, where the confined to the deconfined phase involves an intermediate state where the gauge group $SU(N_c)$ is spontaneously broken into a deconfined $SU(M)$ subgroup and a confined $SU(N_c - M)$ sector.

The present work presents a study of the confined/deconfined phase transition by analysing the Polyakov Loop under the light of the thermofractal approach~\cite{Deppman2016}. The hierarchical structure of thermofractals is present in Yang-Mills fields, where the parameter $q$, associated with the Tsallis Statistics~\cite{Tsallis2023}, can be calculated in terms of the fundamental and adjoint representations parameters. The thermofractal structure indices a fractal geometry of the momentum space~\cite{Deppman2018}.

The GWW model describes eigenvalue interactions as a Coulomb gas on a circle, where repulsion is governed by the Vandermonde determinant derived from the Haar measure. In the large-$N_{c}$ limit, this leads to a hard gap at $\theta=\pi$, causing a sharp "knee" in the $\langle L \rangle$ curve. We argue that this smoothing is a mathematical necessity of a fractal momentum space. In $SU(N)$ gauge theories, the Polyakov loop is defined as the normalised trace of the thermal holonomy. The integration over the gauge group utilises the invariant Haar measure. For the eigenvalues $\{e^{i\theta_j}\}$, the weight is given by the square of the Vandermonde determinant. This gives rise to the repulsive Vandermonde Potential
\begin{equation}
 V(\theta)=-\sum_{j< k}\ln \Big| e^{i\theta}-e^{i\theta'} \Big|^2 \,.
\end{equation}
In the saddle-point, the derivative $dV(\theta)/d\theta$ results in the right-hand side of the central equation in the context of the matrix models, namely, the saddle-point equation:
\begin{equation}
 \frac{2}{g^2} \sin \theta= {\cal P} \int_{-\pi}^{\pi} \rho\big((\theta') \cot \big(\frac{\theta - \theta'}{2} \big) d \theta' \,, \label{eq:saddle-point}
\end{equation}
where $g$ is the coupling of the interacting fields, and $\theta$ and $\theta'$ are the eigenvalues in the compactified circle. The right-hand side is the repulsive force resulting from the Vandermonde Potential. In the continuum limit, the force between two eigenvalues at a relative angle $\theta$ is $F(\theta) =dV(\theta)/d \theta \propto 1/\theta$. For $\theta \rightarrow 0$, the singularity represents the topological stiffness of the $SU(N)$ vacuum in the standard GWW limit.

Hanada's work provides a compelling macroscopic picture. In this work, we study how the thermofractal approach offers a microscopic derivation of the continuous growth of the deconfined ratio degrees of freedom. In this paper, we demonstrate that the topological smoothing necessary for partial deconfinement is a natural consequence of the fractal momentum space structure inherent to thermofractals. By matching the thermofractal index $q$ to the QCD $\beta$-function, we derive unique analytical expressions that describe the continuous transition between these sectors. Unlike phenomenological additions in Polyakov-Nambu-Jona-Lasinio models, our $q$-deformations arise directly from QCD's running coupling.

\section{Thermofractal Scaling and $q(N_{f})$ Scaling}

Thermofractals feature self-similar structures leading to non-extensive statistics. The degree of non-extensivity, index $q$, is matched to the one-loop QCD $\beta$-function. 
\begin{equation}
    \frac{1}{q-1}=\frac{11N_{c}-2N_{f}}{3} \,. \label{eq:qNcNf}
\end{equation}
In $SU(3)$, this relation between $q$, $N_c$ and $N_f$ provides $q=8/7 \approx 1.141$ for the QCD $N_f=6$, a value that has been thoroughly confirmed by momentum distribution analyses in multi-particle production with RICH and LHC experimental data~\cite{Wong2015}. The relation in Eq.~(\ref{eq:qNcNf}) stems from a non-perturbative running-coupling~\cite{Deppman2020}.

In (2+1)-lattice calculations, $N_f=3$ gives $q_l=10/9 \approx 1.111$. This value has been confirmed in studies of quark dynamics using (2+1)-QCD calculations~\cite{Walton2000}.  For $N_f=0$ (gluonic phase), it results $q_g=12/11 \approx 1.091$, and represents a stage where gluons are partially deconfined while quarks remain confined, allowing for sub-critical degrees of freedom.

To study the Centre Symmetry Break in the context of the matrix models, we need to introduce the thermofractal running-coupling, $g(p) \propto e_q(\varepsilon/\Lambda)$, where $\Lambda$ is a scale variable and $\varepsilon$ is the partonic energy. The $q$-exponential function is $ e_q(x)=[1+(q-1)x]^{-1/(q-1)}$. A naive approach would be the substitution of $g$ in Eq.~(\ref{eq:saddle-point}) by the thermofractal; such an approach does not fully capture the hierarchical self-similar structure of thermofractals. The thermal fluctuations of eigenvalues at different levels of the hierarchical structure induce a fractal structure in the eigenvalue circle the is responsible for softening the phase transition.

An effective form for introducing the fractal structure is to recur to the q-deformed derivatives, a method already tested in the study of dynamics in fractal spaces~\cite{Deppman2023,Megas2024}. To account for the fractal vacuum, we apply the $q$-deformed derivative
\begin{equation}
 \mathcal{D}_{q}f(x)=x^{q-1}\frac{d}{dx}f(x) \,.
\end{equation}
To introduce the fractal structure of the momentum space, we replace the standard derivative with the $q$-deformed operator $\mathcal{D}_q$. The modified force $F_q(\theta)$ is obtained by applying $\mathcal{D}_q$ to the $q$-logarithm of the chordal distance:
\begin{equation}
    F_q(\theta) = \mathcal{D}_q \ln_q \left[ 2 \sin\left( \frac{\theta}{2} \right) \right] = \theta^{q-1} \frac{d}{d\theta} \left( \frac{[2 \sin(\theta/2)]^{1-q} - 1}{1-q} \right)
\end{equation}
In the short-distance limit ($\theta \to 0$), the interaction scales as $F_q(\theta) \approx 1/\theta $, keeping the same stiffness of the GWW model.
Crucially, however, the global topology of the potential is shifted, and the phase-transition results smeared.

Consider the $SU(N)$ effective action in the presence of fractal fluctuations. The steady-state eigenvalue density $\rho(\theta)$ is determined by the balance between the modified potential $V_q(\theta)$ and the thermal noise $\xi(T)$ characterised by the thermofractal variance $\sigma^2_{TF}(T)$. 
The steady-state saddle-point equation for the eigenvalue density $\rho(\theta)$ is given by the balance of forces:
\begin{equation}
    \mathcal{D}_q V(\theta) = \int \rho(\theta') F_q(\theta - \theta') d\theta' \,.
\end{equation}
According to non-extensive statistical mechanics, the equilibrium distribution that extremizes the $q$-entropy for a given potential $V(\theta)$ is the $q$-exponential
\begin{equation}
    \rho(\theta)=\frac{1}{\mathcal{N}(q)}\left[1-\frac{q-1}{\sigma^{2}(T)}(1-\cos \theta)\right]^{-\frac{1}{q-1}}
\end{equation}
where $\mathcal{N}(q)$ is the normalization constant. Unlike GWW, $\rho(\theta)$ remains finite for all $\theta \in [-\pi, \pi]$ due to its power-law tails ($\rho(\pi)>0$), representing deconfined fluctuations coexisting with a confined background. The saddle-point equation gives the density 
\begin{equation}
    \rho(\theta) \approx \frac{1}{\mathcal{N}} \left[ 1 - \frac{q-1}{2\sigma^2(T)} \theta^2 \right]^{-\frac{1}{q-1}} \label{eq:themofractaldensity}
\end{equation}
where $\sigma(T)$ is the density width.

The temperature dependence of the distribution is governed by the variance $\sigma^2(T)$, which must preserve the thermofractal energy/temperature fluctuation, resulting
\begin{equation}
\sigma^2(T) = \sigma_0^2 \left[ 1 + (q-1) \frac{T - T_c}{\Lambda } \right]^{\frac{1}{q-1}} \,.
\end{equation}
The parameter $\Lambda$ calibrates the thermofractal scaling to the Lattie-QCD scale.

While the eigenvalue density in Eq.~(\ref{eq:themofractaldensity}) integrates the thermofractal hierarchical structure into the phase-transition, additional topological considerations are needed. Note that the Polyakov loop is a 1D temporal probe, while in the deconfined phase the field $A_0$ undergoes fluctuations governed by the fractal measure $d\mu_q$. The global average involves a spatial integration over $d^3x$:
\begin{equation}
\langle L \rangle = \int [dA] e^{-S_q[A]} \text{Tr} \mathbf{L} \approx \exp\left( -\int d^3x \chi_q(\vec{x}) \right)
\end{equation}
where $\chi_q$ is the local susceptibility of the fractal vacuum.
In terms of the eigenvalue density, the Polyakov loop is the first moment of the density, given by
\begin{equation}
    \langle L(T)\rangle=\frac{\int_{-\pi}^{\pi}\rho(\theta)\,\cos \theta d\theta}{\int_{-\pi}^{\pi}\rho(\theta) d\theta}
\end{equation}
Since the Polyakov loop is localised in space but extended in time, its coupling to the 3D fractal bath results in a suppression exponent proportional to the spatial embedding dimension $d=3$. To account for the 3D spread of the thermofractal fluctuations, the projected distribution width must include the full spatial dimensionality, i.e., $1 - \langle L \rangle \propto \left[ \sigma^2(T) \right]^{-d} = \left[ \sigma^2(T) \right]^{-3}$.
Substituting the field-theoretic stiffness into the geometric projection: $1 - \langle L \rangle \sim (T^7)^{-3} = T^{-21}$.

The physical significance of $\sigma^2(T)$ extends beyond a mere statistical width.  The growth of $\sigma^2(T)$ can be identified as a microscopic proxy for the expansion of the $SU(M)$ deconfined core~\cite{Hanada2024}. While the GWW model relies on a rigid potential to force a third-order jump, the thermofractal variance represents the thermal kinetic energy within fractal clusters that allows eigenvalues to gradually leak into the confined region $[-\pi, \pi]$, smoothing the transition into a crossover. 

\section{The gluonic phase}

This section provides a parameter-free explanation for the short tails observed in the Polyakov Loop below $T_c$. In the GWW framework, the confined phase is characterised by eigenvalues distributed across the entire circle to minimise repulsion. In our model, this corresponds to the regime where the variance $\sigma^2(T)$ is small, making the distribution $\rho(\theta)$ highly localised at $\theta = 0$. Using the gluonic index $q_g = 12/11$, the integration yields the scaling
\begin{equation}
\langle L \rangle_{conf} \approx 1 - \frac{\langle \theta^2 \rangle}{2} \propto [\sigma^2(T)]^{\frac{1}{q_g-1}} = [\sigma^2(T)]^{11}\,.
\end{equation}

In the gluonic phase, $N_f=0$ and the index $q_g = 12/11$, therefore the exponent $1/(q_g - 1)=11$, yielding the algebraic growth $\langle L \rangle \propto [1 + (q_g - 1)\frac{T-T_c}{\Lambda}]^{11}$.
Substituting the temperature dependence $\sigma^2(T) \propto [1 + (q_g-1)\frac{T-T_c}{\Lambda}]$ from Eq. (9), results
\begin{equation}
    \langle L\rangle_{conf} = L_c \left[ 1 + (q_g-1)\frac{T-T_c}{\Lambda} \right]^{\frac{1}{q_g-1}}
\end{equation}
With $(q_g-1)^{-1} = 11$, the loop exhibits a "soft" algebraic growth rather than the hard zero predicted by the GWW model.

In the standard GWW model, the confined phase is characterised by a uniform eigenvalue distribution necessitated by the rigid $1/\theta$ Vandermonde repulsion, leading to a vanishing Polyakov loop. However, within the thermofractal framework, the topological stiffness is smoothed by the $q$-deformed interaction $F_q \sim 1/\theta$. Under the light of the partial deconfinement theory proposed by Hanada et al. \cite{Hanada2024}, we interpret the thermofractal density in the confined regime ($T < T_c$) not as a global state of the vacuum, but as the emergence of a deconfined $SU(M)$ subgroup. 

\begin{figure}[htbp]
    \centering
    \includegraphics[width=0.5\columnwidth]{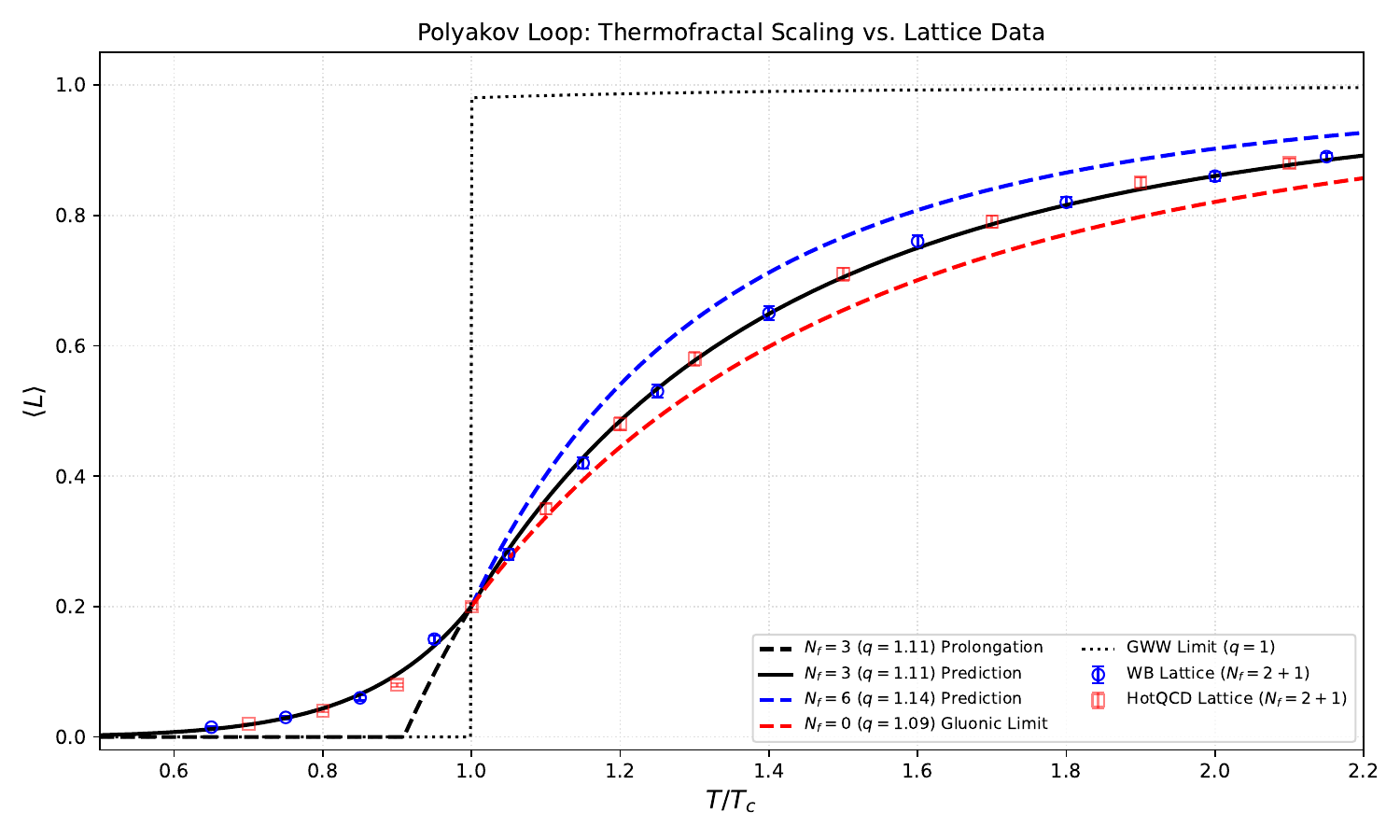} 
    \caption{Temperature dependence of the Polyakov loop $\langle L \rangle$ in $SU(3)$ gauge theory. 
    The blue squares represent $(2+1)$-flavor lattice QCD data~\cite{Borsnyi2010}. 
    The solid black line denotes the Two-Phase Thermofractal model, which incorporates a gluonic 
    phase ($q=12/11$) below $T_c$ and a quark phase ($q=10/9$) above $T_c$, assuming $N_f=3$. 
    The dotted black line represents the rigid Gross-Witten-Wadia (GWW) transition ($q=1$), 
    which exhibits a sharp third-order discontinuity at $T_c$. 
    The dashed black line indicates the pure quark phase ($q=10/9$) extrapolated to temperatures below $T_c$, highlighting the necessity of the gluonic phase at low temperatures. The dashed blue line indicates the QCD result ($N_f=6$), and the dashed red line shows an extrapolation of the pure gluonic phase to $T>T_c$. All cases collapse to the gluonic regime for $T<T_c$.
    The best fit parameters are: $\Lambda=0.141 \pm 0.002$ and $L_c=0.201 \pm 0.005$. 
    The reduced $\chi^2$ is 1.12. Cross-validation with HotQCD data~\cite{Bazavov2024} 
    (red symbols, not used in calibration) yields similar agreement, confirming model independence.}
    \label{fig:polyakov_comparison}
\end{figure}

The low-temperature regime $T < T_c$ is characterised by a state of {\it gluonic self-organised criticality}, aligning with instanton-liquid models where sub-critical fluctuations persist~\cite{Shuryak1988}. In this phase, the $SU(3)$ vacuum is dominated by gluonic fluctuations ($N_f=0$) with a thermofractal index $q_g = 12/11$.  Physically, this represents the emergence of self-similar deconfined clusters within a globally confined background. The scaling of the Polyakov loop in this region is governed by the $q$-exponential stability of the gauge field configurations. The concentration of eigenvalues near $\theta=0$ represents this partially deconfined core. As the variance $\sigma^2(T)$ is small in this regime, the distribution takes the form of a $q$-Gaussian, where the Polyakov loop identifies with the partial deconfinement.

This refinement clarifies that the non-zero values of $\langle L \rangle$ below $T_c$ are not due to global deconfinement, but to the fractal distribution of energy among gluonic degrees of freedom. The power-law exponent of 11 is a direct consequence of the $SU(3)$ $\beta$-function matching, providing a parameter-free link between the vacuum's topological tail and the fundamental coupling of the theory.

The inclusion of the spatial dimension $d=3$ in the deconfined exponent reflects a fundamental topological shift in the gauge field. In the confined regime ($T < T_c$), the thermofractal fluctuations are restricted to the one-dimensional geometry of the colour-flux tubes (strings) connecting the colour sources. Upon crossing $T_c$, these string-like constraints dissolve, allowing the fractal measure to occupy the full three-dimensional volume of the Quark-Gluon Plasma. This dimensional crossover explains why the gluon field behaves effectively as a 1D object until the transition point.

\section{Analysis and Discussion}

The theoretical approach presented in the previous sections allows for an analytical description of the Polyakov Loop in the phase-transition between confined and deconfined regimes. This allowed for a comparison with lattice-QCD data by using a global parameter $\Lambda$ to calibrate the lattice scale, and the parameter $L_c$ to fix the gluonic phase, allowing a continuous behaviour of $\langle L \rangle(T)$ at $T=T_c$, fixing $N_f=3$ for compatibility with the (2+1) lattice data. The parameters were calibrated to $\Lambda=0.141 \pm 0.002$ and $L_c=0.201 \pm 0.005$, with a reduced chi-square of $\chi^2=1.12$. Only the data from Borsanyi et al.~\cite{Borsnyi2010} was used in the calibration, while the HotQCD data~\cite{Bazavov2024} is used for a cross-check of the theoretical approach.

The results are reported in Fig.~\ref{fig:polyakov_comparison}, where it is possible to observe that the theoretical curve for $N_f=3$ accurately describes the data in the entire range of temperatures. Note that the continuation of a pure $N_f=3$ regime cannot completely describe the data below $T_c$. Only after the introduction of the gluonic field can the full lattice data be accurately described in the full range of temperatures. This aspect shows that the gluonic regime is crucial for an accurate description of the lattice data. While our two-phase model captures the crossover, finite-volume effects in lattice data may slightly bias low-T tails~\cite{Cossu2009}, which probably can be accounted for by a recalibration of the parameter $L_c$.

The combination of gluonic field and hierarchical deconfinement introduced by the thermofratal structure provides the correct crossover obtained in the lattice calculations, being able to reproduce the lattice data from Ref.~\cite{} without adjustable free-parameters. This is possible because of two aspects: the thermofractal structure introduces a hierarchically spread noise that manifests during the phase transition as a partial deconfinement, softening the transition. The correct inclusion of the geometric aspects of the physical process on both sides of the critical temperature, using a 3D system for the deconfined phase and a 1D gluon-string for the pre-deconfinement stage, is crucial. In addition, it adheres to the physical picture of the phase transition that is present in many models, like quark-fragmentation and jet-production~\cite{Rapp1999,DEramo2013}.

The results obtained here show that the thermofractal approach provides a microscopic mechanism for partial deconfinement, introducing hierarchical fluctuations that smear the phase transition. The variance $\sigma^2(T)$, which we use to smooth the eigenvalue density, acts as the statistical proxy for the size of the partially deconfined sector. As $\sigma^2(T)$ grows according to Eq. (9), the $SU(M)$ subgroup expands continuously, mapping the GWW third-order jump into the thermofractal crossover. In doing so, it unveils aspects of the phase transitions that were not obviously observed in calculations by other methods, such as the sub-critical fluctuations from a gluon-string, and the three-dimensional character of the deconfined regime.

The physical significance of $\sigma^2(T)$ lies in the determination of the quantum fluctuations at different hierarchical levels of the thermofractal. Effectively, it works as a smoothing parameter for the eigenvalue density $\rho(\theta)$. In the limit $N_f \rightarrow \infty$, the transition becomes sharper, and the results approach the GWW result. In fact, this situation corresponds to $q \rightarrow 1$, when the thermofractal formalism recovers the GWW, and the lack of the temperature-dependent scaling results in a rigid potential that forces a hard gap. In our model, $\sigma^2(T)$ represents the thermal noise within the fractal clusters; as $T$ increases, this noise overcomes the Vandermonde repulsion, allowing eigenvalues to occupy the entire colour circle $[-\pi, \pi]$ without a non-analytic jump. This mechanism reconciles the rigid matrix model topology with the continuous crossover observed in $(2+1)$-flavour lattice QCD.

The thermofractal approach presents a microscopic mechanism for Hanada's partial deconfinement model. The hierarchical fluctuations can be seen as fractal clusters of colored degrees of freedom breaking the Vandermonde barrier. These clusters represent the partially deconfined degrees of freedom that soften the phase transition. Our variance $\sigma^2(T) \propto T^{1/(q-1)}$ quantifies the continuous M-growth in $SU(M)×SU(N_c-M)$ breaking, offering testable predictions for large-$N_c$ extrapolations. In this way, the approach presented here bridges the GWW results and the Hanada's model by offering a continuous mechanism that leads from the sharp transition in the case $q=1$ to smoother transistions for $N_f \rightarrow \infty$, correctly reproducing the (2+1)-lattice data when $N_f=3$.

Furthermore, the transition from $N_f=0$ to $N_f=3$ or $N_f=6$ embodies the chiral phase transition. The gluonic stage corresponds to the strong condensate of the quark field in models like the Polyakov-Nambu-Jona-Lasinio (PNJL), where quarks are massive. As the temperature increases, the condensate vanishes, and the quarks' degrees of freedom are liberated. The results with thermofractals indicate that the chiral phase transition is embedded in the centre symmetry phase transition, and offer new ways for exploring the PNJL by introducing the thermofractal structure in those models. In this concern, the main finding in the thermofractal approach is the 3D-dimension manifestation through the Polyakov Loop that happens exactly at the critical temperature for the chiral symmetry break,

As a final cross-check of the approach introduced here, we explore a critical requirement for any model of the deconfining transition: the recovery of the perturbative QCD limit at high temperatures. We examine the asymptotic behaviour of the Polyakov loop in the limit where $T \gg T_c$. As $T$ increases, the temperature-dependent variance $\sigma^2(T)$ grows following the power-law $\sigma^2(T) \propto T^{1/(q-1)}$.

In this limit, the term $(q-1)(1-\cos\theta)/\sigma^2(T)$ in the eigenvalue density $\rho(\theta)$ becomes negligible for all $\theta \in [-\pi, \pi]$. Consequently, the density distribution approaches the uniform limit:
\begin{equation}
    \lim_{T \to \infty} \rho(\theta) = \frac{1}{2\pi}
\end{equation}
Under this uniform distribution, the eigenvalues are spread evenly across the colour circle, representing a state of maximal colour disorder. Substituting this into the Polyakov loop integral, we find:
\begin{equation}
    \lim_{T \to \infty} \langle L(T) \rangle = \int_{-\pi}^{\pi} \frac{1}{2\pi} \cos\theta \, d\theta = 1
\end{equation}
This mathematical limit corresponds to the physical state of full deconfinement, where the screening of the colour potential is complete, and the free energy of a static quark source vanishes. 

Furthermore, the thermofractal framework ensures that the approach to this limit is governed by the $q$-index derived from the QCD $\beta$-function. Specifically, the deviation from unity scales as:
\begin{equation}
    1 - \langle L(T) \rangle \sim \left( \frac{\Lambda}{T} \right)^{\frac{3}{q-1}}
\end{equation}
For $q_q = 8/7$, this yields a $T^{-21}$ decay of the non-perturbative corrections, which is significantly faster than the corrections typically found in standard polynomial models~\cite{Kajantie2003}, suggesting thermofractals accelerate perturbative recovery This rapid convergence highlights that the "topological smoothing" introduced by the fractal measure is a phenomenon concentrated near the critical region, leaving the high-temperature Quark-Gluon Plasma in its expected perturbative state.

\section{Conclusion}

We have demonstrated that the smooth crossover in $SU(3)$ deconfinement is a mathematical necessity of a fractal momentum space. By matching the thermofractal index $q$ to the QCD $\beta$-function, we provided analytical scaling laws that accurately reproduce lattice QCD data with a reduced $\chi^2 \approx 1.12$. 

Crucially, our results provide a thermofractal foundation for the partial deconfinement theory of Hanada et al. \cite{Hanada2024}. The "topological smoothing" observed in our model is the effective manifestation of the continuous growth of deconfined $SU(M)$ clusters, governed by $q$-exponential statistics rather than the rigid GWW measures. This reconciliation between matrix model topology and non-extensive thermodynamics offers a unified description of the QCD vacuum across the deconfining transition.

Future extensions could apply thermofractals to chiral transitions through PNJL models~\cite{Fukushima2004} or heavy-ion observables, bridging lattice QCD with experiment. The identification of a distinct gluonic stage below $T_c$ provides testable signatures for experimental heavy-ion physics~\cite{DEramo2013}. Specifically, the topological crossover from 1D string-like fluctuations to 3D deconfined volumes should manifest as a characteristic shift in thermal photon emissivity and a specific $T^{11}$ scaling in the low-mass dilepton production rates~\cite{Brodsky2008}, offering a window into the sub-critical self-organisation of the QCD vacuum.

\section{Acknowledgements}
This work was funded by the Conselho Nacional de Desenvolvimento Científico e Tecnológico (CNPq-Brazil), grant number 306093/2022-7,  and FAPESP, grant number 2024/01533-7.

\bibliographystyle{apsrev4-2}
\bibliography{ThermofractalonPolyakovLoop.bib} 

@article{Gross1980,
  title = {Possible third-order phase transition in the large-<mml:math xmlns:mml="http://www.w3.org/1998/Math/MathML" display="inline"><mml:mi>N</mml:mi></mml:math>lattice gauge theory},
  volume = {21},
  ISSN = {0556-2821},
  url = {http://dx.doi.org/10.1103/PhysRevD.21.446},
  DOI = {10.1103/physrevd.21.446},
  number = {2},
  journal = {Physical Review D},
  publisher = {American Physical Society (APS)},
  author = {Gross,  David J. and Witten,  Edward},
  year = {1980},
  month = jan,
  pages = {446–453}
}

@article{Wadia1980,
  title = {N = ∞ phase transition in a class of exactly soluble model lattice gauge theories},
  volume = {93},
  ISSN = {0370-2693},
  url = {http://dx.doi.org/10.1016/0370-2693(80)90353-6},
  DOI = {10.1016/0370-2693(80)90353-6},
  number = {4},
  journal = {Physics Letters B},
  publisher = {Elsevier BV},
  author = {Wadia,  Spenta R.},
  year = {1980},
  month = jun,
  pages = {403–410}
}

@article{Douglas1993,
  title = {Large N phase transition in continuum QCD2},
  volume = {319},
  ISSN = {0370-2693},
  url = {http://dx.doi.org/10.1016/0370-2693(93)90806-S},
  DOI = {10.1016/0370-2693(93)90806-s},
  number = {1–3},
  journal = {Physics Letters B},
  publisher = {Elsevier BV},
  author = {Douglas,  Michael R. and Kazakov,  Vladimir A.},
  year = {1993},
  month = dec,
  pages = {219–230}
}

@article{Bazavov2024,
  title = {Charm degrees of freedom in hot matter from lattice QCD},
  volume = {850},
  ISSN = {0370-2693},
  url = {http://dx.doi.org/10.1016/j.physletb.2024.138520},
  DOI = {10.1016/j.physletb.2024.138520},
  journal = {Physics Letters B},
  publisher = {Elsevier BV},
  author = {Bazavov,  A. and Bollweg,  D. and Kaczmarek,  O. and Karsch,  F. and Mukherjee,  Swagato and Petreczky,  P. and Schmidt,  C. and Sharma,  Sipaz},
  year = {2024},
  month = mar,
  pages = {138520}
}

@article{Hanada2024,
  title = {On Thermal Transition in QCD},
  volume = {2024},
  ISSN = {2050-3911},
  url = {http://dx.doi.org/10.1093/ptep/ptae033},
  DOI = {10.1093/ptep/ptae033},
  number = {4},
  journal = {Progress of Theoretical and Experimental Physics},
  publisher = {Oxford University Press (OUP)},
  author = {Hanada,  Masanori and Watanabe,  Hiromasa},
  year = {2024},
  month = feb 
}

@article{Hanada2024b,
  title = {A New Perspective on Thermal Transition in QCD},
  volume = {2024},
  ISSN = {2050-3911},
  url = {http://dx.doi.org/10.1093/ptep/ptae044},
  DOI = {10.1093/ptep/ptae044},
  number = {4},
  journal = {Progress of Theoretical and Experimental Physics},
  publisher = {Oxford University Press (OUP)},
  author = {Hanada,  Masanori and Ohata,  Hiroki and Shimada,  Hidehiko and Watanabe,  Hiromasa},
  year = {2024},
  month = mar 
}

@article{Watanabe2021,
  title = {Partial deconfinement at strong coupling on the lattice},
  volume = {2021},
  ISSN = {1029-8479},
  url = {http://dx.doi.org/10.1007/JHEP02(2021)004},
  DOI = {10.1007/jhep02(2021)004},
  number = {2},
  journal = {Journal of High Energy Physics},
  publisher = {Springer Science and Business Media LLC},
  author = {Watanabe,  Hiromasa and Bergner,  Georg and Bodendorfer,  Norbert and Funai,  Shotaro Shiba and Hanada,  Masanori and Rinaldi,  Enrico and Sch\"{a}fer,  Andreas and Vranas,  Pavlos},
  year = {2021},
  month = feb 
}

@article{Hanada2019,
  title = {Partial Deconfinement},
  volume = {2019},
  ISSN = {1029-8479},
  url = {http://dx.doi.org/10.1007/JHEP03(2019)145},
  DOI = {10.1007/jhep03(2019)145},
  number = {3},
  journal = {Journal of High Energy Physics},
  publisher = {Springer Science and Business Media LLC},
  author = {Hanada,  Masanori and Ishiki,  Goro and Watanabe,  Hiromasa},
  year = {2019},
  month = mar 
}

@article{nsal2008,
  title = {Abelian Duality,  Confinement,  and Chiral-Symmetry Breaking in a SU(2) QCD-Like Theory},
  volume = {100},
  ISSN = {1079-7114},
  url = {http://dx.doi.org/10.1103/PhysRevLett.100.032005},
  DOI = {10.1103/physrevlett.100.032005},
  number = {3},
  journal = {Physical Review Letters},
  publisher = {American Physical Society (APS)},
  author = {\"{U}nsal,  Mithat},
  year = {2008},
  month = jan 
}

@article{Borsnyi2010,
  title = {Is there still any T c mystery in lattice QCD? Results with physical masses in the continuum limit III},
  volume = {2010},
  ISSN = {1029-8479},
  url = {http://dx.doi.org/10.1007/JHEP09(2010)073},
  DOI = {10.1007/jhep09(2010)073},
  number = {9},
  journal = {Journal of High Energy Physics},
  publisher = {Springer Science and Business Media LLC},
  author = {Borsányi,  Szabolcs and Fodor,  Zoltán and Hoelbling,  Christian and Katz,  Sándor D. and Krieg,  Stefan and Ratti,  Claudia and Szabó,  Kálmán K.},
  year = {2010},
  month = sep 
}

@article{Cossu2009,
  title = {Finite size phase transitions in QCD with adjoint fermions},
  volume = {2009},
  ISSN = {1029-8479},
  url = {http://dx.doi.org/10.1088/1126-6708/2009/07/048},
  DOI = {10.1088/1126-6708/2009/07/048},
  number = {07},
  journal = {Journal of High Energy Physics},
  publisher = {Springer Science and Business Media LLC},
  author = {Cossu,  Guido and D’Elia,  Massimo},
  year = {2009},
  month = jul,
  pages = {048–048}
}

@article{Deppman2016,
  title = {Thermodynamics with fractal structure,  Tsallis statistics,  and hadrons},
  volume = {93},
  ISSN = {2470-0029},
  url = {http://dx.doi.org/10.1103/PhysRevD.93.054001},
  DOI = {10.1103/physrevd.93.054001},
  number = {5},
  journal = {Physical Review D},
  publisher = {American Physical Society (APS)},
  author = {Deppman,  A.},
  year = {2016},
  month = mar 
}

@article{Deppman2020,
  title = {Fractal Structures of Yang–Mills Fields and Non-Extensive Statistics: Applications to High Energy Physics},
  volume = {2},
  ISSN = {2624-8174},
  url = {http://dx.doi.org/10.3390/physics2030026},
  DOI = {10.3390/physics2030026},
  number = {3},
  journal = {Physics},
  publisher = {MDPI AG},
  author = {Deppman,  Airton and Megías,  Eugenio and P. Menezes,  Débora P.},
  year = {2020},
  month = sep,
  pages = {455–480}
}

@book{Tsallis2023,
  title = {Introduction to Nonextensive Statistical Mechanics: Approaching a Complex World},
  ISBN = {9783030795696},
  url = {http://dx.doi.org/10.1007/978-3-030-79569-6},
  DOI = {10.1007/978-3-030-79569-6},
  publisher = {Springer International Publishing},
  author = {Tsallis,  Constantino},
  year = {2023}
}

@article{Deppman2018,
  title = {Fractal Structure and Non-Extensive Statistics},
  volume = {20},
  ISSN = {1099-4300},
  url = {http://dx.doi.org/10.3390/e20090633},
  DOI = {10.3390/e20090633},
  number = {9},
  journal = {Entropy},
  publisher = {MDPI AG},
  author = {Deppman,  Airton and Frederico,  Tobias and Megías,  Eugenio and Menezes,  Debora P.},
  year = {2018},
  month = aug,
  pages = {633}
}

@article{Megas2024,
  title = {Dynamics in fractal spaces},
  volume = {848},
  ISSN = {0370-2693},
  url = {http://dx.doi.org/10.1016/j.physletb.2023.138370},
  DOI = {10.1016/j.physletb.2023.138370},
  journal = {Physics Letters B},
  publisher = {Elsevier BV},
  author = {Megías,  Eugenio and Khalili Golmankhaneh,  Alireza and Deppman,  Airton},
  year = {2024},
  month = jan,
  pages = {138370}
}

@article{Deppman2023,
  title = {Fractal Derivatives,  Fractional Derivatives and q-Deformed Calculus},
  volume = {25},
  ISSN = {1099-4300},
  url = {http://dx.doi.org/10.3390/e25071008},
  DOI = {10.3390/e25071008},
  number = {7},
  journal = {Entropy},
  publisher = {MDPI AG},
  author = {Deppman,  Airton and Megías,  Eugenio and Pasechnik,  Roman},
  year = {2023},
  month = jun,
  pages = {1008}
}

@article{Wong2015,
  title = {From QCD-based hard-scattering to nonextensive statistical mechanical descriptions of transverse momentum spectra in high-energy<mml:math xmlns:mml="http://www.w3.org/1998/Math/MathML" display="inline"><mml:mi>p</mml:mi><mml:mi>p</mml:mi></mml:math>and<mml:math xmlns:mml="http://www.w3.org/1998/Math/MathML" display="inline"><mml:mrow><mml:mi>p</mml:mi><mml:mover accent="true"><mml:mrow><mml:mi>p</mml:mi></mml:mrow><mml:mrow><mml:mo accent="true" stretchy="false">¯</mml:mo></mml:mrow></mml:mover></mml:mrow></mml:math>collisions},
  volume = {91},
  ISSN = {1550-2368},
  url = {http://dx.doi.org/10.1103/PhysRevD.91.114027},
  DOI = {10.1103/physrevd.91.114027},
  number = {11},
  journal = {Physical Review D},
  publisher = {American Physical Society (APS)},
  author = {Wong,  Cheuk-Yin and Wilk,  Grzegorz and Cirto,  Leonardo J. L. and Tsallis,  Constantino},
  year = {2015},
  month = jun 
}

@article{Walton2000,
  title = {Equilibrium Distribution of Heavy Quarks in Fokker-Planck Dynamics},
  volume = {84},
  ISSN = {1079-7114},
  url = {http://dx.doi.org/10.1103/PhysRevLett.84.31},
  DOI = {10.1103/physrevlett.84.31},
  number = {1},
  journal = {Physical Review Letters},
  publisher = {American Physical Society (APS)},
  author = {Walton,  D. Brian and Rafelski,  Johann},
  year = {2000},
  month = jan,
  pages = {31–34}
}

@article{Kajantie2003,
  title = {Pressure of hot QCD up to<mml:math xmlns:mml="http://www.w3.org/1998/Math/MathML" display="inline"><mml:mrow><mml:msup><mml:mrow><mml:mi>g</mml:mi></mml:mrow><mml:mrow><mml:mn>6</mml:mn></mml:mrow></mml:msup></mml:mrow><mml:mi mathvariant="normal">ln</mml:mi><mml:mn/><mml:mo>(</mml:mo><mml:mn>1</mml:mn><mml:mo>/</mml:mo><mml:mi>g</mml:mi><mml:mo>)</mml:mo><mml:mn/></mml:math>},
  volume = {67},
  ISSN = {1089-4918},
  url = {http://dx.doi.org/10.1103/PhysRevD.67.105008},
  DOI = {10.1103/physrevd.67.105008},
  number = {10},
  journal = {Physical Review D},
  publisher = {American Physical Society (APS)},
  author = {Kajantie,  K. and Laine,  M. and Rummukainen,  K. and Schr\"{o}der,  Y.},
  year = {2003},
  month = may 
}

@article{Fukushima2004,
  title = {Chiral effective model with the Polyakov loop},
  volume = {591},
  ISSN = {0370-2693},
  url = {http://dx.doi.org/10.1016/j.physletb.2004.04.027},
  DOI = {10.1016/j.physletb.2004.04.027},
  number = {3–4},
  journal = {Physics Letters B},
  publisher = {Elsevier BV},
  author = {Fukushima,  Kenji},
  year = {2004},
  month = jul,
  pages = {277–284}
}

@article{DEramo2013,
  title = {Momentum broadening in weakly coupled quark-gluon plasma (with a view to finding the quasiparticles within liquid quark-gluon plasma)},
  volume = {2013},
  ISSN = {1029-8479},
  url = {http://dx.doi.org/10.1007/JHEP05(2013)031},
  DOI = {10.1007/jhep05(2013)031},
  number = {5},
  journal = {Journal of High Energy Physics},
  publisher = {Springer Science and Business Media LLC},
  author = {D’Eramo,  Francesco and Lekaveckas,  Mindaugas and Liu,  Hong and Rajagopal,  Krishna},
  year = {2013},
  month = may 
}

@article{Brodsky2008,
  title = {Maximum wavelength of confined quarks and gluons and properties of quantum chromodynamics},
  volume = {666},
  ISSN = {0370-2693},
  url = {http://dx.doi.org/10.1016/j.physletb.2008.06.054},
  DOI = {10.1016/j.physletb.2008.06.054},
  number = {1},
  journal = {Physics Letters B},
  publisher = {Elsevier BV},
  author = {Brodsky,  Stanley J. and Shrock,  Robert},
  year = {2008},
  month = aug,
  pages = {95–99}
}

@article{Rapp1999,
  title = {Low-mass dileptons at the CERN-SpS: evidence for chiral restoration?},
  volume = {6},
  ISSN = {1434-601X},
  url = {http://dx.doi.org/10.1007/s100500050364},
  DOI = {10.1007/s100500050364},
  number = {4},
  journal = {The European Physical Journal A},
  publisher = {Springer Science and Business Media LLC},
  author = {Rapp,  R. and Wambach,  J.},
  year = {1999},
  month = dec,
  pages = {415–420}
}

@book{Shuryak1988,
  title = {The QCD Vacuum,  Hadrons and Superdense Matter},
  ISBN = {9789812799425},
  ISSN = {1793-1436},
  url = {http://dx.doi.org/10.1142/0161},
  DOI = {10.1142/0161},
  journal = {World Scientific Lecture Notes in Physics},
  publisher = {WORLD SCIENTIFIC},
  author = {Shuryak,  E V},
  year = {1988},
  month = mar 
}

% \begin{thebibliography}{99}
% \bibitem{1} D. J. Gross and E. Witten, Phys. Rev. D 21, 446 (1980).
% \bibitem{2} M. Hanada et al., Phys. Rev. D 103, 106005 (2021).
% \bibitem{3} A. Deppman, Phys. Rev. D 93, 054001 (2016).
% \bibitem{4} S. Borsanyi et al., JHEP 09, 073 (2010).
% \bibitem{5} A. Deppman et al., Physics 2, 455 (2020).
% \end{thebibliography}

\end{document}